\documentclass[twoside]{article}

\usepackage{filecontents}

\usepackage{blindtext} 

\usepackage[sc]{mathpazo} 
\usepackage[T1]{fontenc} 
\usepackage{microtype} 

\usepackage[english]{babel} 

\usepackage[hmarginratio=1:1,top=32mm,columnsep=20pt]{geometry} 
\usepackage[hang, small,labelfont=bf,up,textfont=it,up]{caption} 
\usepackage{booktabs} 

\usepackage{lettrine} 

\usepackage{amsmath}
\usepackage{enumitem}
\usepackage{array}
\usepackage{tabularx}

\usepackage{graphics}
\usepackage{graphicx}
\usepackage{colortbl}
\usepackage{multirow}

\usepackage{algorithm}
\usepackage{algorithmic}

\usepackage{enumitem} 
\setlist[itemize]{noitemsep} 

\usepackage{abstract} 

\usepackage{titlesec} 
\renewcommand\thesection{\Roman{section}} 
\renewcommand\thesubsection{\roman{subsection}} 
\titleformat{\section}[block]{\large\scshape\centering}{\thesection.}{1em}{} 
\titleformat{\subsection}[block]{\large}{\thesubsection.}{1em}{} 

\usepackage{fancyhdr} 
\usepackage{titling} 

\usepackage{hyperref} 

\usepackage{fancyhdr}

\usepackage{amsthm}

\newtheorem{definition}{Definition}

\pretitle{\begin{center}\Huge\bfseries} 
\posttitle{\end{center}} 

\title{
 \LARGE{ Geometry-Based Optimization of One-Way Quantum Computation Measurement Patterns\thanks{This paper is an extended version of the paper presented at ICEE 2016~\cite{icee2016}.}}} 
\author{%
\small\textsc{Maryam Eslamy} \\ 
\footnotesize Department of Computer Engineering and Information Technology, Amirkabir University of Technology, Tehran, Iran. \\ 
\footnotesize \href{mailto:maryameslamy@aut.ac.ir}{maryameslamy@aut.ac.ir}
\and 
\small\textsc{Mahboobeh Houshmand} \\[1ex] 
\footnotesize Department of Computer Engineering, Mashhad Branch, Islamic Azad University, Mashhad, Iran \\ 
\footnotesize \href{houshmand@mshdiau.ac.ir}{houshmand@mshdiau.ac.ir} 
\and 
\small\textsc{Morteza Saheb Zamani} \\[1ex] 
\footnotesize Department of Computer Engineering and Information Technology, Amirkabir University of Technology, Tehran, Iran.\\ 
\footnotesize \href{szamani@aut.ac.ir}{szamani@aut.ac.ir} 
\and 
\small\textsc{Mehdi Sedighi} \\[1ex] 
\footnotesize Department of Computer Engineering and Information Technology, Amirkabir University of Technology, Tehran, Iran.\\ 
\footnotesize \href{msedighi@aut.ac.ir}{msedighi@aut.ac.ir} 
}
\date{} 
\renewcommand{\maketitlehookd}{%
\begin{abstract}
{
In one-way quantum computation (1WQC) model, an initial highly entangled state called a
graph state is used to perform universal quantum computations by a sequence of adaptive
single-qubit measurements and post-measurement Pauli-$X$ and Pauli-$Z$ corrections. The needed
computations are organized as measurement patterns, or simply patterns, in the 1WQC model. The entanglement operations in a pattern can be shown by a graph which together with the set of its input and output qubits is called the geometry of the pattern. Since a one-way quantum computation pattern is
based on quantum measurements, which are fundamentally nondeterministic evolutions, there must be conditions over geometries to guarantee determinism.
Causal flow is a sufficient and generalized flow (gflow) is a necessary and sufficient condition over geometries to identify a dependency structure for the measurement sequences in order to achieve determinism. Previously, three optimization methods have been proposed to simplify 1WQC patterns which are called standardization, signal shifting and Pauli simplification. These optimizations can be performed using measurement calculus formalism by rewriting rules.
However, maintaining and searching these rules in the library can be complicated with respect to implementation. Moreover, serial execution of these rules is time consuming due to executing many ineffective commutation rules. To overcome this problem, in this paper, a new scheme is proposed to perform optimization techniques on patterns with flow or gflow only based on their geometries instead of using rewriting rules. Furthermore, the proposed scheme obtains the maximally delayed gflow order for geometries with flow. It is shown that the time complexity of the proposed approach is improved over the previous ones.
}{}{}
\end{abstract}
}

\begin{document}

\maketitle
\onecolumn


\section{Introduction}
Quantum computers which use quantum mechanical phenomena are different from digital electronic computers based on transistors. They are advantageous over the classical computers for solving certain problems such as integer factorization~\cite{shor-1997-26} and database search~\cite{Grover} much more quickly.

One-way quantum computation (1WQC)~\cite{browne},~\cite{PhysRevLett.86} is a more recent model for quantum computation in which a specific highly entangled state called a graph state allows for universal quantum computation by single-qubit measurements and post-measurement Pauli-$X$ and Pauli-$Z$ corrections. The computations are driven by irreversible projective measurements and hence, the model is called ``one-way".

1WQC has attracted researchers' interests as it benefits from two important concepts in quantum mechanics, namely entanglement and measurement. This powerful framework introduces a different quantum computation model from the well-known quantum gate array model which is more similar to classic model of computations by networks of gates. It is also believed to have easier implementations in different quantum technologies~\cite{mbqc,mbqctwo}.

The needed computations in this model are organized as measurement patterns, or simply patterns. The entanglement operations in a pattern can be represented in a graph which together with the set of its input and output qubits is called the geometry of the pattern. Since a pattern is
based on quantum measurements, which are fundamentally nondeterministic evolutions, there should be conditions over geometries to guarantee determinism. Causal flow~\cite{flow} is a sufficient condition and generalized flow (gflow)~\cite{gflow} is a sufficient and necessary one over geometries for identifying a dependency structure for measurement sequences in order to obtain determinism. In 1WQC model~\cite{ACM.54}, three different optimization techniques called standardization, signal shifting and Pauli simplification, which can be done using a set of rewrite rules have been proposed~\cite{ACM.54,silva2013}.

However in these approaches, there should be a library for storing rewrite rules and automatically applying them is time consuming due to using many ineffective commutation rules which are performed on input pattern separately. To overcome this problem, the key result of this study is to provide an automatic approach to simultaneously perform these optimization techniques on patterns with flow or gflow\footnote{In the rest of this paper, whenever we refer to open graphs with gflow, the ones are considered that have gflow but not flow.} only based on their geometries instead of using rewriting rules. This results in lower time complexity than the previous technique~\cite{extended}. Moreover, our method can also calculate \emph{the maximally delayed gflow} order for geometries with flow.

The remainder of this paper is as follows. The background material is presented in the next section. Section~\ref{sec:related} reviews the related work. In Section~\ref{sec:proApproach}, the proposed approach is explained. The proof and analysis of the proposed approach are described in Section~\ref{sec:corectness} and finally, Section~\ref{sec:con} concludes the paper.

\section{Preliminaries}
\label{sec:pre}
In this section, some basic concepts and notations used in the rest of the paper are introduced
and explained.
\subsection{Quantum Circuit Model}
\label{sec:pre1}

Quantum bits or \emph{qubits} are quantum analogues of classical bits. A qubit is a two-level quantum system whose state is represented by a unit vector in a two-dimensional Hilbert space, $\mathcal{H}_2$, for which an orthonormal basis, denoted by $\{$$\left\vert 0\right\rangle$, $\left\vert 1\right\rangle$$\}$, has been fixed. Unlike classical bits, qubits can be a superposition of $\left\vert 0\right\rangle$ and $\left\vert 1\right\rangle$ like $a\left\vert 0\right\rangle+b\left\vert 1\right\rangle$  where $a$ and $b$ are complex numbers such that
$|a|^2 + |b|^2 = 1$.
If the qubit is measured in the $\{$$\left\vert 0\right\rangle$, $\left\vert 1\right\rangle$$\}$ basis, the classic outcome of 0 is observed with the probability of $|a|^2$ and the classic outcome of 1 is observed with the probability of $|b|^2$. If 0 is observed, the state of the qubit after the measurement collapses to $|0\rangle$ and otherwise, it collapses to $|1\rangle$.

\emph{Entanglement} is a quantum mechanical phenomenon that plays a key role in many of the applications of quantum computation and quantum information.
A multi-qubit quantum state $\left\vert \psi \right\rangle$ is said to be entangled if it cannot be written as
the tensor product $\left\vert \psi \right\rangle=\left\vert \phi_1 \right\rangle \otimes \left\vert \phi_2\right \rangle $ of two pure states. For example, the EPR pair~\cite{Nielsen} shown below is an entangled quantum state:
\[\left\vert \Phi  \right\rangle=(\left\vert 00 \right\rangle+\left\vert 11 \right\rangle)/\surd{2}\]

\emph{Quantum-circuit model} is a well-known model of quantum computations~\cite{Deutsch}, based on the unitary evolution of qubits by networks of gates. Every quantum gate is a linear transformation represented by a unitary matrix, defined on an $n$-qubit Hilbert
space. A matrix $U$ is \emph{unitary} if $UU^{\dagger} = I$, where $U^{\dagger}$ is the conjugate transpose of $U$. Some useful single-qubit gates are the elements of the Pauli set which are defined as follows:
\[\sigma_0=
I=%
\begin{bmatrix}
1 & 0\\
0 & 1
\end{bmatrix},
\sigma_1=
X=%
\begin{bmatrix}
0 & 1\\
1 & 0
\end{bmatrix},
\sigma_2=
Y=%
\begin{bmatrix}
0 & -i\\
i & 0
\end{bmatrix},
\sigma_3=
Z=%
\begin{bmatrix}
1 & 0\\
0 & -1
\end{bmatrix}\]

Hadamard, $H$ is another known single-qubit gate defined as:
\[
H=%
\frac{1}{\sqrt{2}}\begin{bmatrix}
1 & 1\\
1 & -1
\end{bmatrix}
\]

If $U$ is a gate that operates on a single qubit, then controlled-\emph{U} gate operates on two qubits, i.e., control and target qubits, and \emph{U} is applied to the target qubit if the control qubit is $\left\vert 1\right\rangle$ and leaves it unchanged otherwise. For example, controlled-\emph{X} (CNOT) gate performs the \emph{X} operator on the target qubit if the control qubit is $\left\vert 1\right\rangle$. Otherwise, the target qubit remains unchanged.

The matrix representation of CNOT gate is
\[
\text{CNOT}=%
\begin{bmatrix}
1 & 0& 0&0\\
0 & 1&0&0\\
0&0&0&1\\
0&0&1&0
\end{bmatrix}\]
Similarly, the matrix representation of controlled-\emph{Z} (CZ) gate is
\[
\text{CZ}=%
\begin{bmatrix}
1&0&0&0\\
0&1&0&0\\
0&0&1&0\\
0&0&0&-1
\end{bmatrix}\]

\subsection{Description of 1WQC model}
\label{sec:pre2}
Measurement patterns, or simply patterns, represent computations in the 1WQC model. A pattern is defined as \emph{P = (V, I, O, A)}, where \emph{V} is the set of qubits, \emph{I} $ \subseteq $ \emph{V} and \emph{O} $ \subseteq $ \emph{V} are two possibly overlapping sets representing the pattern inputs and outputs respectively and \emph{A} is a finite set of operations which act on \emph{V} and is called \emph{command sequence}. The set \emph{V} is called \emph{computation space}.
Using the notations of~\cite{ACM.54}, the following operations can be defined:
\begin{enumerate}
\item 1-qubit auxiliary \emph{preparation} $N_v$ prepares a qubit \emph{v} $ \in $ \emph{V} in the ($\left\vert 0 \right\rangle+\left\vert 1 \right\rangle)/\surd{2}$ state,
\item 2-qubit \emph{entanglement operation} $E_{uv}$ performs a CZ operation on qubits \emph{u,v} $ \in$ \emph{V},
\item 1-qubit \emph{correction operations}  $X_v$ and $Z_v$ apply Pauli \emph{X} and \emph{Z} corrections on qubit \emph{v}, and
\item 1-qubit \emph{measurement operation} $M_v^\alpha$ measures the qubit \emph{v} in the orthonormal basis of:\\
$$\left\vert +_\alpha\right\rangle =\frac{1}{\surd{2}}(\left\vert 0 \right\rangle+e^{i\alpha}\left\vert 1 \right\rangle)\: (with\: outcome\:0)$$
$$\left\vert -_\alpha\right\rangle =\frac{1}{\surd{2}}(\left\vert 0 \right\rangle-e^{i\alpha}\left\vert 1 \right\rangle)\: (with\: outcome\:1)$$
\end{enumerate}
where $\alpha \in [0,2\pi]$ is called the angle of measurement.
The outcome of a measurement on a qubit \emph{v} is called $s_v$, which is referred to as \emph{classical feedforward} of the measurement results. Measurement outcomes can be added modulo 2 and are called \emph{signals}.
A measurement can depend on the other ones through two signals of \emph{s} and \emph{t}:\\
\begin{equation}
\label{eq:edgham}
^t[M_i^\alpha]^s \equiv M_i^\alpha X_i^s Z_i^t \equiv M_i^{((-1)^s+t\pi)}
\end{equation}
The one-qubit Pauli correction commands, $X_v^s$ and $Z_v^s$, apply the Pauli \emph{X} and \emph{Z} gates to the qubit \emph{v} if the signal \emph{s=1} and do nothing if \emph{s=0}:\\
\centerline{$X_v^1=X_v$}\\
\centerline{$Z_v^1=Z_v$}\\
\centerline{$X_v^0=Z_v^0=I_v$}\\
In this paper, it is assumed that all of the non-input qubits are prepared and hence the preparation commands on these qubits are omitted.

The patterns of $J(\alpha)$ and CZ gates are presented as examples.
$J(\alpha)$ plays an important role in 1WQC and is defined as follows:
\[J(\alpha )=\frac{1}{{\sqrt 2 }}
\begin{bmatrix}
1 & {e^{i\alpha } }\\
1 & { - e^{i\alpha } }
\end{bmatrix}\]
The following pattern realizes the $J(\alpha)$ gate:
\[
J(\alpha) = X_2^{s_1 } M_1^{ - \alpha } E_{12}
\]
where $\{1,2\}$ is the set of qubits, $\{1\}$ is the set of input qubits and $\{2\}$ is the set of output qubits.\\
In the rest of the paper, whenever the angle $\alpha$ is not important, the $J(\alpha)$ gate is simply referred to as a $J$ gate.

The following pattern implements CZ gate:
\[\mathrm{CZ} = E_{12}\]
where both $\{1,2\}$ are input and output qubits.

Different criteria may be used to evaluate a pattern. The size and the quantum computation depth of patterns are among the most considered criteria. The size of a pattern refers to the number of qubits involved in it. The quantum computation depth of a pattern or just quantum depth is the maximum number of levels of operations for the execution of the pattern due to the dependencies of measurement and correction commands. For example, the quantum depth of the standard pattern $P = \{ \{ 1,\,2,\,...,5\} ,\,\{ 1\} ,\,\{ 5\} ,\,X_5^{s_2  + s_4 } Z_5^{s_1  + s_3 } M_4^0 [M_3^\alpha  ]^{s_2 } [M_2^\theta  ]^{s_1 } M_1^\beta  E_{12345} \}$ is 4 due to the dependencies of the qubits $1 \to 2 \to 3 \to 5$ where each arrow shows the dependency of a qubit (e.g., 2) on the other (e.g., 1).

\subsubsection{Optimization in 1WQC}\label{sec:opt}
In~\cite{ACM.54}, some optimization techniques were presented which will be used later in this paper. These techniques are introduced in the following.
\paragraph{Standardization}\hspace{0pt}\\
A pattern is said to be in the standard form if all of the entanglement operations appear first, followed by all of the measurement operations and the correction commands at the end of the command sequence.
In other words, a measurement pattern \emph{P = (V, I, O, A)} in the standard form can be written as \emph{CME} where \emph{C}
represents the operators which perform all of the \emph{X} and \emph{Z}
corrections, \emph{M} is the operator attributed to all of the
measurement commands and \emph{E} represents the operators
that perform all of the entanglement commands.
To move the entanglement commands to the beginning of the command sequence, the following rewrite rules can be used:

\begin{eqnarray}
E_{ij} X_i^s  \Rightarrow X_i^s Z_i^s E_{ij}\label{eq:e1}\\
E_{ij} Z_i^s  \Rightarrow Z_i^s E_{ij}\label{eq:e2}\\
E_{ij} A_{\mathop k\limits^ \to  }  \Rightarrow A_{\mathop k\limits^ \to  } E_{ij}\label{eq:e3}
\end{eqnarray}
where \emph{A} is an arbitrary command and ${\mathop k\limits^ \to  }$ represents the qubits acted on by \emph{A} which do not contain \emph{i} and \emph{j}. By using the following commutativity rules, the correction commands can be moved to the end of the pattern:
\begin{eqnarray}
 A_{\mathop k\limits^ \to  } X_i^s  \Rightarrow X_i^s A_{\mathop k\limits^ \to  } \label{eq:c1} \\
 A_{\mathop k\limits^ \to  } Z_i^s  \Rightarrow Z_i^s A_{\mathop k\limits^ \to  }\label{eq:c2}
\end{eqnarray}

The $X_i^s$ and $Z_i^s$ correction commands also commute even when they act
on the same qubit. This holds because the operators \emph{XZ}
and \emph{ZX} differ only by a global phase.
\paragraph{Pauli Simplification}\hspace{0pt}\\
If a measurement angle is 0 ($\frac{\pi }{2}$), the measurement is called a Pauli \emph{X} (Pauli \emph{Y}) measurement. When there is a Pauli \emph{X} measurement on a qubit, the \emph{X} correction dependencies from the qubit can be removed by using Equation \ref{eq:removex}.
\begin{equation}
\label{eq:removex}
M_i^0 X_i^s  = M_i^0
\end{equation}
Similarly, when a Pauli \emph{Y} measurement is performed on a
qubit, the \emph{X} correction dependencies can be changed to \emph{Z} correction dependencies according to Equation \ref{eq:removez}.
\begin{equation}
\label{eq:removez}
M_i^{\frac{\pi }{2}} X_i^s  = M_i^{\frac{\pi }{2}} Z_i^s
\end{equation}
Therefore, there are no \emph{X} corrections on the qubits whose measurement angles are 0 or $\frac{\pi }{2}$.
\paragraph{Signal Shifting}\hspace{0pt}\\
All of the \emph{Z} corrections on the measured qubits can be moved to the end of a pattern. This process is
called signal shifting and the rewrite rules used for it are as the following:

\begin{equation}
^t[ {M_i^\alpha  }]^s \Rightarrow S_i^t \left[{M_i^\alpha  }\right]^s\label{eq:s1}
\end{equation}
\begin{equation}
^t[ {M_j^\alpha  }]^s S_i^r \Rightarrow S_i^{r\,} {}^{t[(r + s_i )/s_i ]}[ {M_j^\alpha  }]^{s[(r + s_i )/s_i ]}\label{eq:s2}
\end{equation}
\begin{equation}
X_j^s S_i^r \Rightarrow S_i^{r\,}X_j^{s[(r + s_i )/s_i ]}\label{eq:s3}
\end{equation}
\begin{equation}
Z_j^s S_i^r \Rightarrow S_i^{r\,}Z_j^{s[(r + s_i )/s_i ]}\label{eq:s4}
\end{equation}
\begin{equation}
S_j^s S_i^r \Rightarrow S_i^r S_j^{s[(r + s_i )/s_i ]}\label{eq:s5}
\end{equation}

where $s[(r + s_i )/s_i ]$  denotes the substitution of $s_i$ with
$r + s_i$ in signal \emph{s} and $S_i^r$ is the signal shifting command which is used to move the signal to the left of 1WQC command sequence.

When an input pattern is simplified by these three optimization techniques, it is called an optimized measurement pattern, or simply optimized pattern.




\subsection{Graph Representation of 1WQC Patterns}\label{sec:Grf_rep}
In order to automatically manipulate 1WQC patterns, each pattern \emph{P} graph in the standard form can be represented as a tuple \emph{G= (V, I, O, E, M, S, T)}~\cite{thesisEinar} where:
\begin{itemize}
\item \emph{V} is the set of vertices in \emph{G} and each vertex represents a qubit,
\item \emph{I} {$\subset$} \emph{V} is the set of input vertices,
\item \emph{O} {$\subset$} \emph{V} is the set of output vertices,
\item \emph{E} is the set of edges in \emph{G},
\item \emph{M} is the set that represents measurement angles for the vertices in \emph{V},
\item \emph{S} is the set of {$S_v$}s. ({$S_v$} is the set of vertices that represents \emph{X} correction dependencies on the vertex \emph{v} $\in$ \emph{V}), and
\item \emph{T} is the set of {$T_v$}s. ({$T_v$} is the set of vertices that represents \emph{Z} correction dependencies on the vertex \emph{v} $\in$ \emph{V}).
\end{itemize}
The set of all vertices, input and output vertices in \emph{G} and those in the pattern \emph{P} are equal. The edges in \emph{G} represent the entanglement commands of \emph{P}. The set \emph{M} contains the measurement angles of \emph{P}.
A vertex \textit{v} is called \emph{X}-dependent (\emph{Z}-dependent) on another vertex \textit{w} if an \emph{X (Z)} correction has to be performed on the qubit \textit{v} which depends on the measurement performed on the qubit \textit{w}.
The \emph{X} correction signal on the qubit \emph{v} in a pattern \emph{P} is calculated by a symmetric difference in \emph{G}:
\begin{equation}
\label{eq:xd}
X_v^{s_1  + s_2  +  \cdot  \cdot  \cdot  + s_m }  \Leftrightarrow S_v  = \left\{ {s_1 } \right\}\Delta \left\{ {s_2 } \right\}\Delta ...\Delta \left\{ {s_m } \right\}\,\,\,\forall v \in \emph{V}
\end{equation}

Similarly, the \emph{Z} correction signal on the qubit \emph{v} in the pattern \emph{P} is calculated by a symmetric difference in \emph{G}:
\begin{equation}
\label{eq:zd}
Z_v^{t_1  + t_2  +  \cdot  \cdot  \cdot  + t_m }  \Leftrightarrow T_v  = \left\{ {t_1 } \right\}\Delta \left\{ {t_2 } \right\}\Delta ...\Delta \left\{ {t_m } \right\}\,\,\,\forall v \in \emph{V}
\end{equation}

A geometry \emph{(G,I,O)} is an entanglement graph $G$, together with subsets $I,O \in V$ representing the sets of input and output vertices of a measurement pattern.

\subsection{Determinism condition in 1WQC}\label{sec:Flow_gflow}

Since a one-way quantum computation pattern is based on quantum measurements, which are fundamentally nondeterministic evolutions.
  The collection of possible measurement outcomes is called the branch of computation in 1WQC model. Because of the probabilistic nature of quantum measurement, there must be conditions over geometries to guarantee determinism. In this paper, the patterns are called deterministic pattern which satisfy three conditions as follows:

  \begin{itemize}
    \item Each branch of computation is obtained with the same probability which is called strong determinism.
    \item Uniform determinism which means that the pattern is deterministic for all values of its measurement angles.
    \item Being deterministic after each single measurement, which is called stepwise determinism.
  \end{itemize}

Flow is a sufficient and generalized flow (gflow) is a sufficient and necessary condition over geometries for the deterministic implementation of measurement patterns over geometries as defined below.

\begin{definition}Flow~\cite{flow}.
An open graph $(G,I,O)$ has a flow iff there exists a map $f: O^C \to I^C$ and a strict partial order $\prec_f$ over $V$ such that all of the following conditions hold for all $i \in O^C$.
\begin{itemize}
  \item $i \prec_f f(i)$,
  \item if $j \in N(f(i))$, then $j=i$ or $i \prec_f j$, where $N(m)$ is the neighbourhood of $m$,
  \item $i \in N(f(i))$.
\end{itemize}
\end{definition}
\begin{definition}Generalised flow (gflow)~\cite{gflow}.
An open graph $(G,I,O)$ has a gflow iff there exists a map $g: O^C \to P^{I^C } \backslash \{ \emptyset \}$  (the set of all non-empty subsets of vertices in $I^C$) and a strict partial order $\prec_g$ over $V$ such that all of the following conditions hold for all $i \in O^C$.
\begin{itemize}
  \item if $j \in g(i)$ then $i \prec_g j$,
  \item if $j \in Odd(g(i))$, then $j=i$ or $i \prec_g j$, where $Odd(K) = \{ k,\left| {N (k) \cap K} \right| = 1\,\,\bmod \,\,2\}$,
  \item $i \in Odd(g(i))$.
\end{itemize}
\end{definition}
In this definition, $Odd(K)$ is the set of vertices which have an odd number of connections with the members in the set $K$. $G(i)$ contains all of the correcting set for the qubit $i$ and it should be noted that flow is a special case of gflow.

As shown in Fig.~\ref{fig:examp2}, the longest path from input to output qubits over those directed edges in the graph defines the depth of the gflow. According to this definition, there is a special type of gflow with minimal depth called maximally delayed gflow~\cite{silva2013,find_gflow}, or simply optimal gflow. In~\cite{find_gflow}, it is proven that there is no gflow of the same graph which is more delayed. In other words, \emph{the maximally delayed gflow} is unique for the input graph with flow. We refer the reader to~\cite{silva2013,find_gflow} to find more details about \emph{the maximally delayed gflow}.

\section{Related work}\label{sec:related}
Raussendorf and Briegel~\cite{Robert} first proved the universality of 1WQC by translating quantum circuits
containing arbitrary single-qubit rotations and CNOT gates to patterns.
A calculus for 1WQC was presented in~\cite{ACM.54}. In that paper, the optimization techniques, standardization, signal shifting and Pauli simplifications were presented. In~\cite{par}, an approach for parallelizing quantum circuits was proposed. This approach takes a circuit solely consisting of CZ and \emph{J} gates and produces the corresponding pattern after performing the mentioned optimizations. Then it translates the optimized 1WQC patterns back to quantum circuits to parallelize the initial quantum circuits. This algorithm was modified in~\cite{silva2013} by introducing a set of rewrite rules to remove the qubits added during the transformation. It leads to a lower depth when an optimized pattern is translated back to a quantum circuit. Moreover, a new theoretical link is shown between maximally delayed generalized flow (the gflow with the minimal depth) and signal shifting. In~\cite{houshi}, by modifying the approach in~\cite{par}, an automatic method was presented to translate quantum circuits consisting of CNOT and any single-qubit gates to optimized
 1WQC patterns.

After flow was introduced in~\cite{flow} as a sufficient condition to implement a deterministic pattern, in~\cite{find_gflow}, an $O(n^2)$ algorithm was proposed to find an optimal flow (flow of minimal depth) in geometries with $n$ vertices. In~\cite{find_flow}, an algorithm of order $O(k^2n)$ has been described to produce a flow in geometries with $n$ vertices and $k$ output vertices. In~\cite{gflow}, a weaker version of flow called generalized flow (gflow) has been introduced to be a sufficient and necessary condition for determinism. In~\cite{find_gflow} a polynomial time algorithm which finds an optimal gflow of an input geometry with $n$ vertices running with $O(n^4)$ has been proposed.

\section{Proposed Approach}
\label{sec:proApproach}

In this section, the proposed approach is explained which takes an arbitrary geometry with flow or gflow as an input and produces an optimized pattern as an output without using rewriting rules. Before starting the main algorithm, we perform a preprocessing phase which determines whether the input geometry has the flow or gflow condition. If the input geometry has flow, it is considered as a geometry with flow. Otherwise, it is dealt with under the case of geometries with gflow (as gflow is a weaker version of flow). The proposed approach slightly differs for geometries with flow or gflow.


The optimized patten is obtained by considering the interactions between the qubits of the underlying input geometry and their neighbors. It obtains the $XList$ and $ZList$ (the lists that consist of all qubits from which a qubit receives
an $X$-correction and a $Z$-correction respectively) for all qubits in the standard pattern after signal shifting and Pauli
simplification. This new method can also calculate \emph{the maximally delayed gflow} order for the input geometry with flow.

The proposed approach uses a graph structure to represent geometries in which each vertex shows a qubit and each edge represents an entanglement operation between two qubits. Moreover, the input and output qubits are determined in sets $I$ and $O$, respectively. In this structure, a qubit is an object and has the attributes as shown in Table~\ref{tab:str}.

\emph{QList} which is filled according to $\prec _f$ obtained by the optimal flow~\cite{find_gflow} or $\prec _g$ obtained by the gflow order over the input geometry consists of all qubits in it. The qubits are arranged in different levels with respect to their orders in the optimal flow or the gflow order and output qubits are put at the end of this list.
\emph{XDependencyList} and \emph{ZDependencyList} are two lists that consist of qubits that can be potentially put in $XList$ and $ZList$ of the qubit at hand, $q$, respectively. The qubits in these two lists are added to the final $XList$ and $ZList$ lists of the qubit $q$ if their \emph{Odd} attribute is TRUE. 
The main method includes four algorithms as explained in the following.

\begin{enumerate}
  \item Algorithm \emph{FindNeighborZQubit} \emph{(QList)}

  This algorithm finds $Z$-\emph{dependency} \emph{neighborhood}~\cite{silva2013} of all qubits in the input geometry. $Z$-\emph{dependency} \emph{neighborhood} is defined as: $N_Z (j)=\left \{k \in O^C |f(k) \in N(j)\setminus {f(j)} \right \}$. It contains the set of qubits from which $j$ receives a $Z$-\emph{correction}. This definition has only been used for geometryies with flow, hence we extend it for geometries with $gflow$. New definition of $Z$-$dependency$ neighborhood set which is used for all geometries with $flow$ or $gflow$ is defined as: $N_Z (j)$=$\left \{ k \in O^C |fg(k) \in N(j) \, where\, k \neq j \right \}$.
  The importance of this new definition is explained in the following sections, when we calculate \emph{XDependencyList} and \emph{ZDependencyList} to determine $XList$ and $ZList$ for each qubit.
   Algorithm~\ref{alg:NZ} shows the pseudo code to create this set.

\begin{algorithm}
\caption {\emph{FindNeighborZlist}(\emph{QList})}
\label{alg:NZ}
\small
\begin{algorithmic}[1]
\FOR{(each $q$ in $QList$)}
\FOR {(each $N$ in \emph{q.NeighborList})}
\STATE add all qubits in $fg^{-1}(N)$ to \emph{q.NeighborZList} except \emph{$fg^{-1}(N)$ = $q$};
\ENDFOR
\ENDFOR
\end{algorithmic}
\end{algorithm}

\begin{table}
\setlength{\arrayrulewidth}{0.3mm}
\setlength{\tabcolsep}{2pt}
\renewcommand{\arraystretch}{1.1}
\caption{Description of qubit attributes}
\label{tab:str}
\scriptsize
\centering
    \begin{tabular} { | p{1.8cm} | p{12cm} |}
    \rowcolor[gray]{0.9}
    \makebox [1.8cm][c]{\textbf{Attribute}} & \makebox [7cm][c]{\textbf{Description}}\\ \hline
     \makebox [2cm][c]{\textbf{\emph{q.ZList}}}
 & This list consists of all qubits from which the qubit $q$ receives a $Z$-\emph{correction}. This list is empty at first.\\ \hline
 \makebox [2cm][c]{\textbf{\emph{q.XList}}}
 & This list consists of all qubits from which the qubit $q$ receives an $X$-\emph{correction}. This list is empty at first.\\ \hline
  \makebox [2cm][c]{\textbf{\emph{q.angle}}}

 & It shows the qubit measurement angle.\\ \hline
  \makebox [2cm][c]{ \textbf{\emph{q.Level}}}
 & It shows the qubit level in \emph{the maximally delayed gflow} order, $\prec_s$, and is calculated for non-output qubits in the input geometry with flow. \\ \hline
 \makebox [2cm][c]{ \textbf{\emph{fg(q)}}}
  & This parameter is filled according to the input geometry with flow or gflow. If the input geometry has the flow condition, this parameter is used to keep the \emph{flow} of each qubit in the geometry. For the input geometry with gflow, it contains the \emph{gflow} of each qubit in the geometry. This parameter is empty for the output qubits in each case. \\ \hline
  \makebox [2cm][c]{ \textbf{\emph{$fg^{-1}(q)$}}}
  & This parameter is filled according to the input geometry with flow or gflow. If the input geometry has the flow condition, this parameter is used to keep the \emph{reverse flow} of each qubit in the geometry. For example, if $f(v)=q$, then $f^{-1}(q)=v$. It is obtained by modifying the algorithm in~\cite{find_gflow} which produces the optimal flow over geometries. Otherwise, if it has the gflow condition, this parameter consists of the \emph{reverse gflow} of each qubit in the geometry. This parameter is empty for the input qubits in each case. \\ \hline
  \makebox [2cm][c]{\textbf{\emph{q.NeighborList}}}
 & This list includes all vertices which are the neighbors of the qubit $q$ in the geometry.\\ \hline
  \makebox [2cm][c]{\textbf{\emph{q.NeighborZLis}}}
 & This list consists of all qubits that are $Z$-\emph{dependency} \emph{neighborhood} of the qubit $q$. It is the output of the function \emph{FindNeighborZQubit}.\\ \hline
  \makebox [2cm][c]{\textbf{\emph{q.Odd}}}
 & This parameter shows whether the number of qubits from which the qubit $q$ receives a $Z$-\emph{correction} (directly or indirectly) is odd or not. It is initialized as FALSE for the first time and is complemented in each loop. \\ \hline
    \end{tabular}
\end{table}

\item Algorithm \emph{FindXList($q$)}

  This algorithm finds all qubits from which $q$ receives an $X$-\emph{correction}. The input to this algorithm is the qubit $q$ and its output is \emph{XList(q)}. This algorithm can also calculate the level of qubit $q$ in \emph{maximally delayed generalised flow order} for the input geometry with optimal flow order. Algorithm~\ref{alg:XL} presents its pseudocode.

  According to the input geometry condition and by using the concept of flow~\cite{flow} or gflow~\cite{gflow}, at first, for each qubit in $fg^{-1}(q)$ set, one qubit is selected and its \emph{Odd} property is complemented. Then, it is added to \emph{XDependencyList}. After that, for each qubit belonging to the \emph{ZList} of the selected qubit, its \emph{Odd} property is complemented. Moreover, it is added to \emph{XDependencyList} if it has not been previously included.

  After these steps, if there is a qubit in \emph{XDependencyList} whose \emph{Odd} property is TRUE, then it is added to \emph{q.Xlist}. This means that the number of $X$-\emph{corrections} from the mentioned qubit to the qubit $q$ is odd. To determine the level of qubit $q$ in \emph{maximally delayed generalised flow order} in the input geometry with flow order, if $q$ is an input qubit, then \emph{q.Level} is set to zero. Otherwise, the maximum qubit level which belongs to the \emph{XDependencyList} is first calculated and then \emph{q.Level} is set to the maximum value of level plus one. After that, all variables are initialized to their default values and \emph{XDependencyList} is cleared. Finally, according to the Pauli simplification and Pauli correction commands~\ref{sec:pre2}, if \emph{q.angle} is equal to $\frac{\pi }{2}$ or 0, then \emph{q.Xlist} is cleared.

\begin{algorithm}[!ht]
\caption {\emph{FindXList}($q$)}
\label{alg:XL}
\small
\begin{algorithmic}[1]
\FOR{(each qubit in $fg^{-1}(q)$)}
\STATE select the \emph{qubit} in this set;
\STATE \emph{qubit.Odd} = ! \emph{qubit.Odd};
\STATE add \emph{qubit} to \emph{XDependencyList};
\FOR {(each \emph{squbit} in \emph{qubit.ZList})}
\IF {(\emph{squbit} was not added to \emph{XDependencyList} beforehand)}
\STATE add \emph{squbit} to \emph{XDependencyList};
\ENDIF
\STATE \emph{squbit.Odd} = ! \emph{squbit.Odd};
\ENDFOR
\ENDFOR
\IF {(\emph{XDependencyList} is empty)}
\STATE \emph{q.Level=0};
\STATE set $\emph{q.Xlist}=\{\}$;
\ELSE
\STATE \emph{max=0};
\FOR {(each \emph{qubit} in \emph{XDependencyList})}
\IF {(\emph{qubit.Odd }== {\bf TRUE})}
\STATE add \emph{qubit} to \emph{q.Xlist};
\STATE \emph{qubit.Odd}={\bf FALSE};
\IF {(\emph{max $<$ qubit.Level})}
\STATE \emph{max=qubit.Level};
\ENDIF
\STATE remove the qubit from \emph{XDependencyList};
\ENDIF
\ENDFOR
\STATE \emph{q.Level=max+1};
\ENDIF
\IF {(\emph{XDependencyList} is not empty)}
\STATE clear it;
\ENDIF
\IF {(\emph{q.angle} is equal to $\frac{\pi }{2}$ or 0)}
\STATE clear \emph{q.Xlist};
\ENDIF
\end{algorithmic}
\end{algorithm}

\item{Algorithm \emph{FindZList($q$)}}

    This algorithm takes the qubit $q$ as an input and its output is the list of qubits for which $Z$-\emph{correction} to $q$ exists. This list is shown by \emph{ZList(q)}. By applying the principle of signal shifting~\cite{ACM.54}, it must check the qubits in \emph{q.NeighborZList}. When the selected qubit angle $q.angle$ is equal to $\frac{\pi }{2}$, it must also check the qubits in $fg^{-1}(q))$ in addition to \emph{q.NeighborZList} according to the principle of Pauli simplification~\cite{ACM.54} converting $X$-\emph{correction} command to $Z$-\emph{correction} one.
   The rest of this algorithm is fundamentally similar to \emph{FindXList(q)}.

   The pseudocode of this procedure is shown in Algorithm~\ref{alg:ZL}.

\begin{algorithm}[!ht]
\caption {\emph{FindZList}($q$)}
\label{alg:ZL}
\small
\begin{algorithmic}[1]
\IF {($q.angle$ is $\frac{\pi }{2}$)}
\FOR {(each \emph{qubit} in \emph{q.NeighborZList} and all qubits in $fg^{-1}($q$))$}
\STATE \emph{qubit.Odd} = !\emph{qubit.Odd};
\STATE add \emph{qubit} to \emph{ZDependencyList};
\FOR {(each \emph{squbit} in \emph{qubit.ZList})}
\IF {(\emph{squbit} was not added to \emph{ZDependencyList} beforehand)}
\STATE add \emph{squbit} to \emph{ZDependencyList};
\ENDIF
\STATE \emph{squbit.Odd} = !\emph{squbit.Odd};
\ENDFOR
\ENDFOR
\ELSE
\FOR {(each \emph{qubit} in \emph{q.NeighborZList})}
\STATE \emph{qubit.Odd} = !\emph{qubit.Odd};
\STATE add \emph{qubit} to \emph{ZDependencyList};
\FOR {(each \emph{squbit} in \emph{qubit.ZList})}
\IF {(\emph{squbit} was not added to \emph{ZDependencyList} beforehand)}
\STATE add \emph{squbit} to \emph{ZDependencyList};
\ENDIF
\STATE \emph{squbit.Odd} = !\emph{squbit.Odd};
\ENDFOR
\ENDFOR
\ENDIF
\IF {(\emph{ZDependencyList} is empty)}
\STATE set $\emph{q.Zlist}=\{\}$;
\ELSE
\FOR {(each \emph{qubit} in \emph{ZDependencyList})}
\IF {(\emph{qubit.Odd} == {\bf TRUE})}
\STATE add \emph{qubit} to \emph{q.Zlist};
\STATE \emph{qubit.Odd} = {\bf FALSE};
\STATE remove the \emph{qubit} from \emph{ZDependencyList};
\ENDIF
\ENDFOR
\ENDIF
\IF {(\emph{ZDependencyList} is not empty)}
\STATE clear it;
\ENDIF
\end{algorithmic}
\end{algorithm}

\begin{figure} [!h]
\centerline{\includegraphics{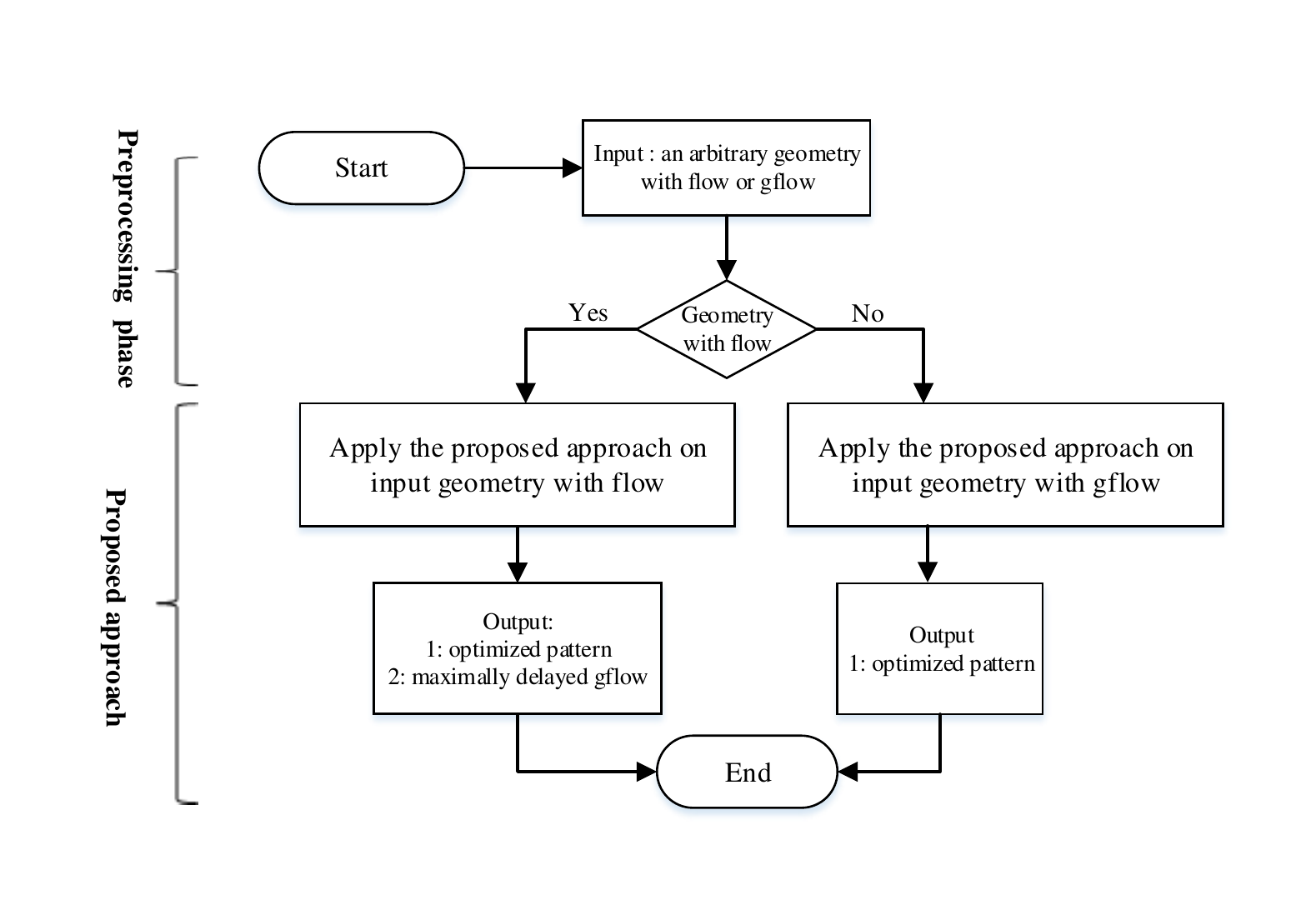}} 
\vspace*{13pt}
\caption{\label{fig:flowchart1}Summary of the proposed approach}
\end{figure}

\item{Algorithm \emph{OptimizitionGeometry(QList)}}

 This algorithm produces the optimized pattern for a given arbitrary geometry with flow or gflow. In this algorithm, initially, \emph{ZList} and \emph{XList} are calculated for all qubits in \emph{QList}.
 The $XList$ as well as measurement commands are printed for all qubits. The $ZList$s are only printed for the output qubits as the $Z$-\emph{corrections} on the non-output qubits are moved to the end of the pattern by applying signal shifting. Its pseudocode is shown in Algorithm~\ref{alg:simpat}.

\begin{algorithm}[!ht]
\caption {\emph{OptimizitionGeometry(QList)}}
\label{alg:simpat}
\small
\begin{algorithmic}[1]
\STATE print entanglement commands as outputs.
\FOR {(each $q$ in \emph{QList})}
\STATE \emph{FindXList($q$)};
\STATE \emph{FindZList($q$)};
\STATE print $X_q^{\emph{q.Xlist}}$;
\STATE print $M_q^{\emph{q.angle}}$;
\IF {$q$ is an output qubit}

\STATE print $Z_q^{\emph{q.Zlist}}$;
\ENDIF
\ENDFOR
\end{algorithmic}
\end{algorithm}
\end{enumerate}

Fig.~\ref{fig:flowchart1} summarizes the proposed approach. Detection of flow or gflow conditions for input geometry is done in the preprocessing phase which is not included in the pseudo code.


%

In the following, we will run this approach on two arbitrary geometries to optimize them. The first geometry has the optimal flow order and the second one has gflow.

%
   \textbf{Example 1}
  In the first example, we apply the proposed approach to a sample geometry with optimal flow order is shown in Fig.~\ref{fig:examp1}. The partial order of this optimal flow is given below.

\begin{center}
  $1 \prec_f 2,4 \prec_f 5,7 \prec_f 8 \prec_f 9$
\end{center}

The step by step results of applying the proposed method are shown in Table~\ref{tbl:flow}.

The output pattern is printed as follows.
\begin{equation*}
\begin{split}
Z_{10}^{s_8  + s_5 + s_4 } X_{10}^{s_7  + s_9 } Z_6^{s_4  + s_7 } X_6^{s_5  + s_2 + s_4} Z_3^{s_1 + s_4  + s_7 } X_3^{s_2}[M_9^{\frac{3 \pi }{9}}]^{s_8 + s_5 + s_4}[M_8^{\frac{\pi }{10}}]^{s_7 }\\
[M_2^{\frac{\pi }{10}}]^{s_1 } M_5^{\frac{\pi }{2}} M_7^{\frac{\pi }{11}}  M_4^{\frac{5 \pi }{9}}M_1^{\frac{\pi }{9}}E_{910} E_{89} E_{68} E_{78} E_{56}E_{45} E_{38} E_{35} E_{23} E_{12}
\end{split}
\end{equation*}

\begin{table}
\setlength{\arrayrulewidth}{0.3mm}
\setlength{\tabcolsep}{2pt}
\renewcommand{\arraystretch}{1.1}
\caption{Step by step running of the proposed method on sample geometry of Fig.~\ref{fig:examp1}}
\label{tbl:flow}
\centering
{
    \begin{tabular}{ | p{1cm} | p{1cm} | p{1cm} | p{2.7cm} | p{2.5cm} | p{2.5cm} |p{1cm} |}
    \rowcolor[gray]{0.9}
    \makebox [1cm][c]{\textbf{Lable}} & \makebox [1cm][c]{\textbf{$f^{-1}(q)$} }& \makebox [1cm][c]{\textbf{angle }}& \makebox [2.7cm][c]{\textbf{NeighborZ }}& \makebox [2.5cm][c]{\textbf{Xlist}}& \makebox [2.5cm][c]{\textbf{Zlist}}& \makebox [1cm][c]{\textbf{Level}}\\ \hline
1& $\oslash$ & $\frac{\pi }{9}$ & $ N_{Z(1)}= \{ \}$ & $X_1=\{ \} $ &$ Z_1=\{ \}$ & 1 \\
2& ${1}$ & $\frac{\pi }{10}$ & $ N_{Z(2)}= \{ \}$ & $X_2=\{ 1 \} $ &$ Z_2=\{ \}$ & 2\\
3& ${2}$ & $-$ & $ N_{Z(3)}= \{ 4,7,1\}$ & $X_3=\{ 2 \} $ & $ Z_3=\{ 4,7,1\}$  & $-$\\
4& $\oslash$ & $\frac{5 \pi }{9}$ & $ N_{Z(4)}= \{ \}$ & $X_4=\{  \} $ & $Z_4=\{  \} $ & 1 \\
5& ${4}$ &  $\frac{\pi }{2}$ &  $ N_{Z(5)}= \{ 2\}$ & $X_5=\{ \} $ &  $ Z_5=\{ 2,4\}$ & 2 \\
6& ${5}$ & $-$ & $ N_{Z(6)}= \{ 4,7\}$ & $X_6=\{ 5,2,4 \}$ & $Z_6=\{ 4,7\}$ & $-$\\
7& $\oslash$ & $\frac{\pi }{11}$ & $ N_{Z(7)}= \{ \}$ & $X_7=\{ \} $ & $ Z_7=\{ \}$ & 1\\
8& ${7}$ & $\frac{\pi }{10}$ & $ N_{Z(8)}= \{ 5,2\}$ & $X_8=\{ 7\} $ & $ Z_8=\{ 5,4\}$ & 2\\
9& ${8}$ & $\frac{3 \pi }{9}$ & $ N_{Z(9)}= \{ 7\}$ & $X_9=\{ 8,5,4\} $ & $ Z_9=\{ 7\}$  & 3\\
10& ${9}$ & $-$ & $ N_{Z(10)}= \{ 8\}$ & $X_{10}=\{9,7 \} $ & $ Z_{10}=\{ 8,5,4\}$ & $-$\\ \hline
    \end{tabular}
    }

\end{table}

\begin{figure} [htbp]
\centerline{\includegraphics{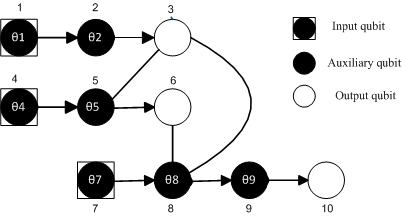}} 
\vspace*{13pt}
\caption{\label{fig:examp1}A sample geometry with flow}
\end{figure}

  \textbf{Example 2}

   In the second example, the new approach is run on a sample geometry with gflow~\cite{silvaF} and shown as Fig.~\ref{fig:examp2}. The partial order of this gflow is given below.

\begin{center}
  $1 \prec 3,5$
\end{center}


  \begin{figure} [htbp]
\centerline{\includegraphics[height=70mm,scale=0.9]{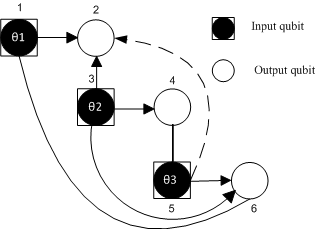}} 
\vspace*{13pt}
\caption{\label{fig:examp2}A sample geometry with gflow~\cite{silvaF}.}
\end{figure}

Algorithm~\ref{alg:simpat} is applied to this geometry with gflow to optimize it. The results are given in Table~\ref{tbl:Gflow_graph}.

\begin{table}
\setlength{\arrayrulewidth}{0.3mm}
\setlength{\tabcolsep}{2pt}
\renewcommand{\arraystretch}{1.1}
\caption{Step by step running of the proposed method on sample geometry of Fig.~\ref{fig:examp2}.}
\label{tbl:Gflow_graph}
\centering
{
    \begin{tabular}{ | p{1cm} | p{1.5cm} | p{1cm} | p{3cm} | p{2.2cm} | p{2.2cm} |p{1cm} |}

    \rowcolor[gray]{0.9}
    \makebox [1cm][c]{\textbf{Lable}} & \makebox [1.5cm][c]{\textbf{$g^{-1}(q)$} }& \makebox [1cm][c]{\textbf{angle}}& \makebox [3cm][c]{\textbf{NeighborZ }}& \makebox [2.2cm][c]{\textbf{Xlist}}& \makebox [2.2cm][c]{\textbf{Zlist}}\\ \hline
1 & $\oslash$ & $\theta_1$ & $ N_{Z(1)}= \{3,5,3,5 \}$ & $X_1=\{ \} $ &$ Z_1=\{ \}$  \\
2 & $\{1,5,3 \}$ & $-$ & $ N_{Z(2)}= \{ \}$ & $X_2=\{5,3\} $ &$ Z_2=\{ \}$\\
3 & $\oslash$ & $\theta_2$ & $ N_{Z(3)}= \{1,5,5\}$ & $X_3=\{  \} $ & $ Z_3=\{1\}$\\
4 & $\{3\}$ & $-$ & $ N_{Z(4)}= \{ \}$ & $X_4=\{1,3\} $ & $Z_4=\{  \} $ \\
5 &  $\oslash$ &  $\theta_3$ &  $ N_{Z(5)}= \{3,3\}$ & $X_5=\{ \} $ &  $ Z_5=\{ \}$\\
6 &  $\{3,5\}$ & $-$ & $ N_{Z(6)}= \{ \}$ & $X_6=\{1,5,3\}$ & $Z_6=\{ \}$ \\ \hline
    \end{tabular}
    }
\end{table}

The output pattern is printed as follows.
\begin{equation*}
\begin{split}
 X_{6}^{s_1  + s_5 + s_3 } X_4^{s_1  + s_3 } X_2^{s_5 + s_3} M_5^{\theta_5} M_3^{\theta_3} M_1^{\theta_1}E_{36} E_{16}E_{56} E_{45} E_{34} E_{23} E_{12}
\end{split}
\end{equation*}

\section {\textbf{Proof and analysis}}\label{sec:corectness}

  In this section, the correctness of the proposed algorithm is proved. Moreover, its time complexity is analysed.

  \subsection{Validation of the proposed algorithm}
 \textbf{Proposition 1.} The proposed approach produces the optimized pattern on an arbitrary geometry with flow or gflow.

 \vspace*{12pt}
\noindent
{\bf Proof:} The proof is done by induction.\\
  a) Base case: The empty pattern, i.e., the pattern with no commands, is optimized. \\
  Proof (a): The base case is trivially correct.\\
  b) Inductive step: Assume that the proposed algorithm generates the optimized pattern for the qubits in the level $k$ of the optimal flow order~\cite{find_gflow} or of the gflow order, $k \geq 1$. Now, it is proved that it produces the optimized pattern for the qubits in the level $k+1$. \\
  Proof (b):

  It is proved that the proposed approach can generate $XList$ and $ZList$ of qubits in the level $k+1$ properly. Consider a qubit $q$ in the level $k+1$. In order to compute the \emph{XList} of $q$, the qubits that can potentially belong to this list, i.e., \emph{XDependencyList} should be considered.
  By the concept of flow~\cite{flow} or gflow~\cite{gflow}, depending on the input geometry, the $X$-correction for $q$ is $p=f^{-1}(q)$ or $p=g^{-1}(q)$ respectively and should be included in \emph{XDependencyList}. By the principal of signal shifting~\cite{ACM.54}, the qubits in the \emph{ZList} of $p$ should also be added to this list.
   Then, the qubits in the \emph{XDependencyList} of $q$ should be included in the \emph{XList} of this qubit if an odd number of them exists. These steps are performed by \emph{FindXList(q)}. As mentioned in~\cite{silva2013}, there is a new connection between \emph{maximally delayed gflow} and signal shifting. Therefore this function can also calculate \emph{maximally delayed gflow order} for the qubits of an arbitrary geometry with flow order. According to the Pauli simplification commands, there is no \emph{XList} set for a qubit with the angle is equal to $\frac{\pi }{2}$ or 0. Therefore this set is cleared for it at the end of this algorithm.

   In order to compute the \emph{ZList} of $q$, the qubits that can potentially belong to this list, i.e., \emph{ZDependencyList} should be created with respect to $q.angle$.
   By using the definition of \emph{NeighborZ}~\cite{silva2013}, if $q.angle$ is not equal to $\frac{\pi }{2}$, the \emph{NeighborZ} of these qubits and their \emph{ZList} qubits are the ones that belong to this list where $ZList$ qubits should be parts of a pattern which are optimized in the level $k$ or less than it. If this angle is equal to $\frac{\pi }{2}$, according to the Pauli simplification commands, the set $fg^{-1}(q)$ and their $ZList$ are included in addition to its \emph{NeighborZ} and their \emph{ZList} qubits to calculate \emph{q.ZList}. According to this technique, the \emph{X}-correction is converted to \emph{Z}-correction for a qubit with the angle equal to $\frac{\pi }{2}$ and then the \emph{Z}-correction is deleted by signal shifting technique. This is true by the inductive condition. The qubits in \emph{ZDependencyList} should be added to the \emph{ZList} of $q$ if an odd number of this exists. These steps are done by~\emph{FindZList(q)}.

As mentioned in~\cite{silva2013}, there is a structural link between signal shifting optimisation and maximally delayed gflow in geometries
with flow. It means when the input and output sizes
of the geometries with flow are equal, signal shifted flow is indeed \emph{the maximally delayed gflow}, or simply optimal gflow. In the proposed
approach, a new procedure is presented that performs signal shifting optimization accurately, so according to the previous theorem
in~\cite{silva2013}, the new method can calculate maximally delayed gflow for geometries with flow correctly.


\begin{table}
\setlength{\arrayrulewidth}{0.3mm}
\setlength{\tabcolsep}{2pt}
\renewcommand{\arraystretch}{1.1}
\caption{Some signal shifted patterns extracted by the proposed approach}
\label{tab:table1}
\footnotesize
\centering
    \begin{tabular}{ | p{0.1cm} | p{2.3cm} | p{8cm}| }
    \rowcolor[gray]{0.9}
    \makebox [0.1cm][c]{\textbf{N}} & \makebox [2.3cm][c]{\textbf{Input circuit} }& \makebox [6cm][c]{\textbf{Optimized pattern }}\\ \hline

   1 & $\mathcal{V} = \frac{{1 + i}}{2}\begin{bmatrix}1 & -i  \\-i & 1  \\\end{bmatrix}$  &  $V:\{1,2,3\}$ $I:\{1\}$ $O:\{3\}$\\
& &$C:X_3^{s_2+s_1 } Z_3^{s_1}$\\
& &$M: M_2^{0} M_1^{ -\frac{\pi }{2}}$\\
& &$E: E_{23} E_{12}$\\
& &\\ \hline

1& $V^{\dagger} = \frac{{1 - i}}{2}\begin{bmatrix}1 & i  \\i & 1  \\\end{bmatrix}$  &$V=\{1,2,3,4,5\}$, $I=\{1\}$, $O=\{5\}$\\
&& $C:X_5^{s_4  + s_2 + s_1 } Z_5^{s_1  + s_3 } M_4^0[{M_3^{ -\pi } }]^{s_2 + s_1 } $\\
& &$M:M_2^{ - \frac{\pi }{2}} M_1^{0}$\\
& &$E:E_{45} E_{34} E_{23} E_{12}$\\
& & \\ \hline

2 & CNOT & $V:\{1,2,3,4\}$ $I:\{1,2\}$ $O:\{1,4\}$\\
& &$C:X_4^{s_3} Z_4^{s_2} Z_1^{s_2}$\\
&&$M:M_3^0 M_2^0$\\
&&$E:E_{34} E_{23} E_{13}$\\
& & \\ \hline

2 &SWAP&$V:\{1,2,3,...,8\}$ $I:\{1,2\}$ $O:\{6,8\}$\\
& &$C:X_8^{s_5  + s_7 } Z_8^{s_1  + s_4 } X_6^{s_3  + s_5 } Z_6^{s_2  + s_4 }$\\
&&$M:M_7^0 M_5^0 M_4^0 M_3^0 M_1^0$\\
& &$E:E_{78} E_{47} E_{56} E_{45} E_{15} E_{34} E_{23} E_{13}$ \\
& & \\ \hline

2 &Bell states circuit~\cite{Nielsen}&$V=\{1,2,3,4,5\}$, $I=\{1,2\}$, $O=\{3,5\}$\\
 & &$C:X_5^{s_1 + s_4}Z_5^{s_2}X_3^{s_1}Z_3^{s_2}$\\
 &&$M:M_4^{0}M_2^{0}M_1^{0}$\\
 &&$E:E_{4,3}E_{2,4,5}E_{1,3}$\\
 & & \\ \hline

3 & Toffoli~\cite{Nielsen}&$V:\{1,2,...,17\}$ $I:\{1,2,3\}$ $O:\{16,17,11\}$\\
& & $C:X_{17}^{s_{12}+s_{14}+s_{15}}Z_{17}^{s_2+s_4+s_6+s_{13}}X_{16}^{s_{15}}Z_{16}^{s_1+s_6+s_8+s_{13}}$\\
& & $X_{11}^{s_4+s_6+s_8+s_{10}}Z_{11}^{s_3+s_5+s_7+s_9}$\\
&&$M:M_{15}^{0}M_{14}^{0}[M_{13}^{\frac{\pi}{4}}]^{s_{12}}M_{12}^{0}[M_{10}^{-\frac{\pi}{4}}]^{s_3+s_5+s_7+s_9}M_9^0$\\
&&$[M_8^{\frac{\pi}{4}}]^{s_3+s_5+s_7}M_7^{0}[M_6^{-\frac{\pi}{4}}]^{s_3+s_5}M_5^{0}[M_4^{\frac{\pi}{4}}]^{s_3}M_3^{0}M_2^{-\frac{\pi}{4}}M_1^{-\frac{\pi}{4}}$\\
& &$E:E_{9,10}E_{19}E_{89}E_{78}E_{27}E_{67}E_{56}E_{15} E_{45}E_{34} E_{23}E_{14,17}E_{14,16}$\\
& &$E_{15,16}E_{1,15}E_{13,14}E_{12,13}E_{1,12}E_{2,12}E_{10,11}$\\
 & & \\ \hline

 2 & CNOT with negative control~\cite{Nielsen}&$V :\{ 1,2,3,...,8\}$ $I :\{1,2\}$ $O: \{8,6\}$\\
& &$C:X_8^{s_3+s_7}Z_8^{s_2+s_4+s_1}X_6^{s_3+s_5}Z_6^{s_2} $\\
& &$M:[M_7^{-\pi}]^{s_1+s_2+s_4}M_5^0M_4^0[M_3^{-\pi}]^{s_1}M_2^0M_1^0$\\
&&$E:E_{78}E_{47}E_{45}E_{25}E_{56}E_{34}E_{13}$ \\
& & \\ \hline

2 &Example 2~\cite{silva2013}&$V=\{1,2,3,...,6\}$, $I=\{1,4\}$, $O=\{3,6\}$\\
& &$C:Z_6^{s_1+s_4}X_6^{s_1+s_5} Z_3^{s_1} X_3^{s_4+s_2}$\\
& &$M:M_5^0 M_2^{\frac{\pi }{2}} M_4^{\theta_4} M_1^{\theta_1}$\\
&&$E:E_{2,5}E_{2,4}E_{456}E_{123}$ \\
& & \\ \hline

3 & Example 1~\cite{silva2013} & $V=\{1,2,3,4,5,6,7,8\}$, $I=\{1,4,7\}$, $O=\{3,6,8\}$\\
& &$C:X_8^{s_7+s_4+s_5} Z_6^{s_1+s_4}X_6^{s_5+s_4+s_2} Z_3^{s_4} X_3^{s_1+s_2+s_4}$\\
 &&$M:[M_5^{\theta_5}]^{s_4+s_1} [M_2^{\theta_2}]^{s_1} M_7^{\theta_7} M_4^{\theta_4} M_1^{\theta_1}$ \\
 &&$E:E_{7,8}E_{6,7}E_{3,7}E_{5,7}E_{3,5}E_{2,5}E_{4,5,6}E_{2,4}E_{123}$ \\
& & \\ \hline

2 & Example 1~\cite{silvaF} & $V=\{1,2,3,4,5\}$, $I=\{1,3,5\}$, $O=\{2,4,5\}$\\
& &$C:X_5^{s_3+s_1} X_4^{s_3+s_1} X_2^{s_1}$\\
 &&$M_3^{\theta_3} M_1^{\theta_1}$ \\
 &&$E:E_{1,5}E_{3,4}E_{3,2}E_{1,4}E_{1,2}$ \\
& & \\ \hline

    \end{tabular}
\end{table}

\subsection{Time complexity analysis}

The time complexity of the algorithms that are used in the proposed method to perform optimization techniques are calculated in the following. The time complexity of each function is calculated for both geometries with flow or gflow separately.

\begin{itemize}
\item \emph{Preprocessing phase}:

An $O(|V|^2)$-algorithm~\cite{find_gflow} is performed to check whether the input geometry has a flow. Otherwise, the input geometry is considered with gflow.

  \item \emph{FindNeighborZQubit(QList)}:

The time complexity of this function for geometries with flow is similar to that of the geometries with gflow.
There are two nested loops in this algorithm. The number of iterations in the outer loop is equal to the number of vertices in the geometry, i.e., $O(|V|)$. The inner loop repeats as many times as the number of neighbors of a qubit which is at most equal to $\Delta (G)$ where $\Delta (G)$ is the largest degree in the graph. Therefore, the total time complexity is equal to:

\begin{center}
  $O(|V|)*O(\Delta (G))=O(|V|\Delta (G))$
\end{center}

  \item \emph{FindXList (q)}:

First, the time complexity is analyzed for geometries with flow and then it is explained for the ones with gflow.

The operations in this algorithm are as follows:

\emph{FindXlist(q)} consists of two nested loops. The outer loop executes ones because there is one qubit belonging to $fg^{-1} (q)$. The reverse flow of each qubit consists of one qubit. The inner loop which is iterated by the number of qubits in \emph{ZDependency(q)}. This list consists of all qubits in the geometry in the worst case. Hence, the time complexity of these nested loops is $O(|V|)$.

There is another loop in this algorithm which has the runtime of $O(|V|)$. The total time complexity is as shown in the following:

\begin{center}
 $O(|V|)+O(|V|)= O(|V|)$
\end{center}

For the geometries with gflow, the number of qubits in $fg^{-1} (q)$ are equal to all qubits belonging to the input geometry in the worst case. Therefore the outer loop takes $O(V)$ time . The inner loop runs in $O(|V|)$ as mentioned for the input geometry with flow. Finally the total runtime is equal to $O(|V|^2)$.

There is another loop in this method which has the runtime of $O(V)$. The total time complexity is as the following:

\begin{center}
$O(V^2)+O(V)= O(V^2)$
\end{center}

\item \emph{FindZlist (q)}:

First this function is analyzed for geometries with flow.
 There are two nested loops for the qubit with angle equal to $\frac{\pi }{2}$. The outer and inner loops are performed in $O(\Delta(G)+1)$ and $O(|V|)$, respectively. As a result, the total time complexity is $O(\Delta(G)|V|)$.

 When the qubit angle is not equal to $\frac{\pi }{2}$, the outer loop has the runtime of $O(\Delta(G))$ and the inner one is performed in $O(|V|)$. Therefore, the total time complexity is $O(\Delta(G)|V|)$. For calculating the \emph{ZList} of output qubits, there is a loop which runs as many times as the number of qubits in \emph{ZDependencyList} and is done in $O(|V|)$ in the worst case. Therefore, the total runtime is as follows:

 \begin{center}
   $O(\Delta(G)|V|)+O(\Delta(G)|V|)+O(V)=O(\Delta(G)|V|)$
 \end{center}

For geometries with gflow, each of the outer and the inner loops are performed in $O(|V|)$. Therefore, the total time complexity is $O(|V|^2))$.

The rest for this case is the same as the previous one and takes $O(V)$ time. The total time is as follows:

 \begin{center}
   $O(|V|^2)+O(|V|^2)+O(V)=O(|V|^2)$
 \end{center}

  \item \emph{OptimizitionGeometry(QList)}:

  Total time complexity of this procedure for geometries with flow is as follows:

\begin{center}
  $O(|V|\Delta(G)|V|)=O(|V|^2\Delta(G))$
\end{center}

 For input graph with gflow, total time complexity of this procedure is shown as follows:

\begin{center}
  $O(|V||V|^2)=O(|V|^3)$
\end{center}

\end{itemize}

The time complexity of the proposed approach is compared to the previous studies in Table~\ref{tbl:compTime}.


\begin{table}[!ht]
\setlength{\arrayrulewidth}{0.3mm}
\setlength{\tabcolsep}{2pt}
\renewcommand{\arraystretch}{1.1}
\caption{Comparison of the proposed approach to the previous studies}
\label{tbl:compTime}
\centering
    \begin{tabular}{ | c | c | c| c| }

    \rowcolor[gray]{0.9}
    \textbf{Algorithm} & \textbf{Input}& \textbf{Output} & \textbf{Time Complexity }\\\hline

 Proposed algorithm & A geometry with flow & An optimized pattern  & $O(|V|^2\Delta(G))$  \\
                          &                     &                       &     \\\cline{2-4}
                          & A geometry with gflow & An optimized pattern & $O(|V|^3)$     \\
                          &                       &                      &              \\\hline
 ~\cite{extended}& A pattern with flow or gflow  & An optimized pattern & $O(|V|^5)$   \\
                      &                               &        &              \\\hline

%
%
%
%
%

    \end{tabular}
\end{table}

As this table shows, the proposed approach for optimizing patterns has a lower time complexity than~\cite{extended}. This method can be applied to valid patterns with flow or gflow. In~\cite{extended} the time complexity is only calculated for performing standardization, so we complete this procedure to find the time complexity for signal shifting and pauli simplification algorithms as well. As mentioned in~\cite{extended}, after applying the standardization technique, we have at most $O(|V|^2)$ entanglement commands at the beginning, $O(|V|)$ measurement commands and finally $O(|V|^3)$ correction commands in the standard pattern. The real complexity of the signal shifting algorithm comes from creating at most $O(|V|)$ signals (Equation~\ref{eq:s1}), each of which has to be commuted by at most $O(|V|^3)$ measurement and correction commands. Therefore, this algorithm has $O(|V|^4)$ time complexity in the worst case. For Pauli simplification, in the worst case, for each measurement command with the angle ($\frac{\pi }{2}$) according to Equation~\ref{eq:removez}, there are $O(|V|)$ $Z$-corrections. Then we can remove them through signal shifting optimization. Hence, the algorithm has a worst case complexity of $O(|V|^4)$ time. To sum up, after performing optimization techniques, the overall time complexity is computed as $O(|V|^5)$.

 If the proposed approach is run on geometries with flow, it takes $O(|V|^2\Delta(G))$. $\Delta(G)$ in a graph is at most of $O(|V|)$ where the worst case is not typical in geometries with flow. Therefore, in the worst case, the proposed approach leads to $O(|V|^3)$ for geometries with flow. Finally, the proposed approach can simplify an arbitrary geometry with flow or gflow with a lower time complexity than the previous study~\cite{extended}. 

One point that needs to be mentioned is that the proposed method, when applied to geometries with flow, can also find \emph{the maximally delayed gflow}. According to Fig.~\ref{fig:flowchart1}, if an input arbitrary geometry has the optimal flow order, then we can also find \emph{the maximally delayed gflow} for it. The time complexity for the first step is $O(|V|^2)$~\cite{find_gflow} and for the second one is $O(|V|^2\Delta(G))$. Therefore, for the whole procedure, it takes $O(|V|^2\Delta(G))$ which has a lower time complexity than~\cite{find_gflow} working on arbitrary geometries. As mentioned before in the worst case it takes $O(|V|^3)$. This is the same as the time complexity in~\cite{silva2013} which can only be applied to geometries with flow to find \emph{the maximally delayed gflow}. Although the time complexity is not improved in this case, the proposed approach also optimizes the input geometry with flow or gflow as well as finding \emph{the maximally delayed gflow} for geometries with flow.

It should be noted that the class of patterns with flow is an interesting class of patterns, as it is universal for quantum
computing and more importantly, the translation from circuits to the patterns in~\cite{par} always leads to a pattern with flow~\cite{par}.

\section {Experimental Results}
\label{sec:exp}
The proposed algorithm was implemented in \emph{C++} on a work station with 4GB RAM and Core 5 Due 2.3GHz CPU. Several examples from the literature, whose
measurement patterns in the optimized form have been manually extracted,
were used and the same results were obtained. Some examples of the optimized patterns extracted by the proposed approach are shown in Table~\ref{tab:table1} where $N$ shows the number of qubits in the input geometry.

\section{Conclusion}\label{sec:con}
In this paper, an algorithm was proposed which takes a geometry with flow or gflow as an input and all of the optimization techniques are applied to it simultaneously without using rewrite rules. These techniques are performed only by checking the neighbors of each vertex in the input geometry. The correctness of the proposed approach was proved and its time complexity analysis showed that it can optimize patterns with flow or gflow with a lower time complexity than the previous approach in~\cite{extended}.

We could also find the order of non-output qubits in the maximally
delayed gflow order for geometries with flow by improving the time complexity compared to
the existing methods in~\cite{find_gflow}. Finding a new connection between two optimal gflow order of a given geometry and signal shifting remains as an interesting open question. If this becomes true then we will conclude that the proposed approach can also calculate an optimal gflow order for any geometry with a time complexity lower than previous approaches.


\bibliographystyle{unsrt}
\bibliography{References}

\begin{thebibliography}{10}

\bibitem{icee2016}
M.~Eslamy, M.~Houshmand, M.~Saheb Zamani, and M.~Sedighi.
\newblock Geometry-based signal shifting of one-way quantum computation
  measurement patterns.
\newblock In {\em 24th Iranian Conference on Electrical Engineering (ICEE
  2016)}, 2016.

\bibitem{shor-1997-26}
P.~W. Shor.
\newblock Polynomial-time algorithms for prime factorization and discrete
  logarithms on a quantum computer.
\newblock {\em SIAM Journal on Computing}, 26:1484--1509, 1997.

\bibitem{Grover}
L.~K. Grover.
\newblock A fast quantum mechanical algorithm for database search.
\newblock {\em ACM Symposium on Theory of Computing}, pages 212--219, 1996.

\bibitem{browne}
D.~Browne and H.~J. Briegel.
\newblock One-way quantum computation - a tutorial introduction.
\newblock 2006.
\newblock http://arxiv.org/abs/quant-ph/0603226.

\bibitem{PhysRevLett.86}
R.~Raussendorf and H.~J. Briegel.
\newblock A one-way quantum computer.
\newblock {\em Physical Review Letters}, 86(22):5188--5191, May 2001.

\bibitem{mbqc}
P.~Walther, K.~J. Resch, T.~Rudolph, and E.~Schenck.
\newblock Experimental one-way quantum computing.
\newblock {\em Nature}, 434:169--176, 2005.

\bibitem{mbqctwo}
H.~J. Briegel, D.~E. Browne, W.~D¨ur, R.~Raussendorf, and M.~Van den Nest.
\newblock Measurement-based quantum computation.
\newblock {\em Nature Physics}.

\bibitem{flow}
V.~Danos and E.~Kashefi.
\newblock Determinism in the one-way model.
\newblock {\em Phys. Rev. A}, 74(5):052310, 2006.

\bibitem{gflow}
D.~Browne, E.~Kashefi, M.~Mhalla, and S.~Perdrix.
\newblock Generalized flow and determinism in measurement-based quantum
  computation.
\newblock {\em New Journal of Phys}, 9, 2007.

\bibitem{ACM.54}
V.~Danos, E.~Kashefi, and P.~Panangaden.
\newblock The measurement calculus.
\newblock {\em Journal of ACM}, 54, 2007.

\bibitem{silva2013}
R.~{Dias da Silva}, E.~Pius, and E.~Kashefi.
\newblock Global quantum circuit optimization.
\newblock {\em Journal of Quantum information and Computation}, 2015.

\bibitem{extended}
V.~Danos, E.~Kashefi, P.~Panangaden, and S.~Perdrix.
\newblock {\em Extended measurement calculus}, pages 235--310.
\newblock Semantic Techniques in Quantum Computation, Cambridge University
  Press, 2009.

\bibitem{Nielsen}
M.~A. Nielsen and I.~L. Chuang.
\newblock {\em Quantum computation and quantum information}.
\newblock Cambridge University Press, 10th anniversary edition edition, 2011.

\bibitem{Deutsch}
D.~Deutsch.
\newblock Quantum computational networks.
\newblock {\em Proc. R. Soc. Lond. A}, 425(1868):73--90, September 1989.

\bibitem{thesisEinar}
E.~Pius.
\newblock Automatic parallelisation of quantum circuits using the measurement
  based quantum computing model.
\newblock Master's thesis, High Performance Computing University of Edinburgh,
  2010.
\newblock
  http://www.epcc.ed.ac.uk/sites/default/files/Dissertations/2009-2010/Einar
  Pius.pdf.

\bibitem{find_gflow}
M.~Mhalla and S.~Perdrix.
\newblock Finding optimal flows efficiently.
\newblock In {\em Proc. of 35th ICALP}, pages 857--868, 2008.

\bibitem{Robert}
R.~Raussendorf.
\newblock {\em Measurement-based quantum computation with cluster states}.
\newblock PhD thesis, Ludwig-Maximilians-Universitat Munich, 2012.

\bibitem{par}
A.~Broadbent and E.~Kashefi.
\newblock Parallelizing quantum circuits.
\newblock {\em Theoretical computer science}, 410(26):2489--2510, June 2009.

\bibitem{houshi}
M.~Houshmand, M.~{Saheb Zamani}, M.~Sedighi, and M.~H. Samavatian.
\newblock Automatic translation of quantum circuits to optimized one-way
  quantum computation patterns.
\newblock {\em Quantum Information Processing}, 13(11), 2014.

\bibitem{find_flow}
N.~Beaudrap.
\newblock Finding ﬂows in the one-way measurement model.
\newblock {\em Phys. Rev. A}.

\bibitem{silvaF}
R.~{Dias da Silva} and E.F. Galvão.
\newblock Compact quantum circuits from one-way quantum computation.
\newblock {\em Physical Review A}, 2013.

\end{thebibliography}

%


\end{document}